\documentclass[10pt]{article} 
\usepackage[utf8]{inputenc}
\usepackage[T1]{fontenc}
\usepackage{amsmath,amssymb,amsfonts}
\usepackage{graphicx}
\usepackage{textcomp}
\usepackage{xcolor}
\usepackage{todonotes}
\usepackage{qcircuit}
\usepackage{nicefrac}
\usepackage{mathtools}
\usepackage{physics}
\usepackage{dcolumn}
\usepackage{bm}
\usepackage{authblk} 

\usepackage{textcomp}

\usepackage{hyperref}
\usepackage[numbers]{natbib}
\usepackage{tikz} 

\usepackage[a4paper, margin=1.0in]{geometry} 

\def\BibTeX{{\rm B\kern-.05em{\sc i\kern-.025em b}\kern-.08em
    T\kern-.1667em\lower.7ex\hbox{E}\kern-.125emX}}
\xyoption{color}
\newcommand{\qwxred}[1][-1]{\ar @[red]@{-} [#1,0]}
\newcommand{\gatered}[1]{*+<.6em>{#1} \POS ="i","i"+UR;"i"+UL **[red]\dir{-};"i"+DL **[red]\dir{-};"i"+DR **[red]\dir{-};"i"+UR **[red]\dir{-},"i" \qw}
\newcommand{\controlred}{*!<0em,.025em>-=-<.2em>{\color{red}\bullet}}
\newcommand{\ctrlred}[1]{\controlred \qwxred[#1] \qw}
\colorlet{darkgreen}{green!60!black}
\colorlet{brightyellow}{yellow!75!red}
\colorlet{orange}{red!50!yellow}
\colorlet{darkblue}{blue!60!black}
\colorlet{darkred}{red!80!black}
\colorlet{greenblue}{green!50!blue}

\def\imag{{\mathrm{i}}}

\title{Performance Evaluations of Signed and Unsigned Noisy Approximate Quantum Fourier Arithmetic}

\author[1,2,3,*]{Robert A. M. Basili}
\author[4,1]{Wenyang Qian}
\author[1,2]{Shiplu Sarker}
\author[1]{Shuo Tang}
\author[3]{Austin Castellino}
\author[3]{Mary Eshaghian-Wilner}
\author[2]{Ashfaq Khokhar}
\author[3]{Glenn Luecke}
\author[1]{James P. Vary}

\affil[1]{\small Department of Physics and Astronomy, Iowa State University, Ames, Iowa, USA}
\affil[2]{Department of Electrical and Computer Engineering, Iowa State University, Ames, Iowa, USA}
\affil[3]{Department of Physics and Mathematics, Iowa State University, Ames, Iowa, USA}
\affil[4]{Instituto Galego de Fisica de Altas Enerxias, Universidade de Santiago de Compostela, Santiago de Compostela, Spain}
\affil[*]{Correspondence: basiliro@iastate.edu}

\begin{document}

\AddToHook{shipout/background}{
    \begin{tikzpicture}[remember picture,overlay]
        \node[anchor=north west, font=\sffamily\bfseries\small, xshift=1cm, yshift=-1cm] at (current page.north west) {Preprint -- arXiv:2112.09349 -- Published in Journal of Supercomputing, 81, 1465 (2025). DOI: 10.1007/s11227-025-07819-1};
    \end{tikzpicture}
}

\date{}

\maketitle

\begin{abstract}
\noindent The Quantum Fourier Transform (QFT) grants competitive advantages, especially in resource usage and circuit approximation, for performing arithmetic operations on quantum computers, and offers a potential route towards a numerical quantum-computational paradigm. In this paper, we utilize efficient techniques to implement QFT-based integer addition and multiplications. These operations are fundamental to various quantum applications including Shor's algorithm, weighted sum optimization problems in data processing and machine learning, and quantum algorithms requiring inner products.
We carry out performance evaluations of these implementations based on IBM's superconducting qubit architecture using different compatible noise models. We isolate the sensitivity of the component quantum circuits on both one-/two-qubit gate error rates, and the number of the arithmetic operands' superposed integer states. We analyze performance, and identify the most effective approximation depths for unsigned quantum addition and quantum multiplication within the given context. We then perform a similar analysis of signed addition and compare to the unsigned results. We observe significant dependency of the optimal approximation depth on the degree of machine noise and the number of superposed states in certain performance regimes.
Finally, we elaborate on the algorithmic challenges - relevant to signed, unsigned, modular and non-modular versions - that could also be applied to current implementations of QFT-based subtraction, division, exponentiation, and their potential tensor extensions. We analyze the performance trends in our results and speculate on possible future developments within this computational paradigm. 
\end{abstract}

\maketitle

\section{Introduction}
%
%
%
%
%
Since its proposal in 1994, the development of Shor's Algorithm \cite{shor1997} for factoring numbers in polynomial time has drawn intense interest in quantum computing, and recent technological leaps have quelled many of the doubts around its practical feasibility. While many challenges must be overcome before fault-tolerant quantum computing becomes remotely possible, the full range of cases where quantum advantage might be achieved remains largely speculative. 

Presently, problem-solving techniques in quantum computing on Noisy Intermediate-Scale Quantum (NISQ) devices are generally considered within the context of a computational basis of qubits acted upon by quantum logic circuits \cite{Preskill2018quantumcomputingin}. In this paradigm, one conventionally maps a space of possible solutions to bit strings, and uses unitary operations (abstracted as quantum logic circuits) to operate on them. The mapping itself may be in some sense arbitrary, but practically must be designed to utilize machine-efficient unitaries. 

However, just as the problem-solving paradigms used in classical computing historically advanced beyond binary representations as resources became more plentiful and robust, the potential roles of quantum and/or quantum-classical hybrid data structures remain largely unknown. What will higher-level programming with fault-tolerant quantum computers look like? What forms of abstraction will be used? While these larger questions extend beyond the scope of this work, we conduct these calculations in part as progenitorial steps towards the potential realization of such a numerical quantum-computational paradigm, and to garner insight on the performance challenges that must be overcome before such a paradigm can be realistically explored. For more immediate purposes, we seek to elucidate practical, near-term limitations of current quantum devices, gauge the effectiveness of a class of approximate circuits quantitatively, and garner insight on what degree of this approximation is optimal for contemporary regimes of machine performance. 

Specifically, we consider quantum arithmetic performed with the Quantum Fourier Transform (QFT) and approximate QFT (AQFT). Quantum arithmetic circuits are the fundamental building blocks of numerous quantum algorithms, and have consequently attracted significant attention. The QFT - a phase-estimation algorithm - is the source of speed-up in Shor's algorithm, and provides several benefits for enabling quantum arithmetic. The approximate QFT improves the algorithm's performance and fidelity on noisy machines by removing gates whose impact on fidelity is less than that of the noise their inclusion imposes. Beyond quantum arithmetic's role in Shor's algorithm, its unique advantage of allowing many operations in parallel presents potential opportunities for a host of problems in data processing, machine learning, and more \cite{kopczyk2018quantum}. 



Although the anticipated utility of quantum arithmetic has given rise to many theoretical implementations of algorithms and provided great incentive to explore their usage, effectively gauging an implementation's performance and identifying sources of error remains a significant obstacle. Barring recent years, considerations of algorithm performance remained fairly limited beyond its theoretical discussion \cite{chakrabarti2008}. Further, despite recent improvements to NISQ hardware, limited access, the small number of qubits, multiple sources of noise, and the relatively large circuit sizes of quantum arithmetic, have made progress in studying performance extremely limited. At the time of this writing, metrics to quantify the novel aspects of that performance specific to quantum arithmetic also remain poorly developed. Ref.~\cite{Methachawalit2020} provides performance results of quantum addition of up to 2-bit integers on the IBM Quantum superconducting-qubit testbed, and observes extremely limited fidelity with early hardware. While not specific to arithmetic, Ref.~\cite{Leymann_2020} provides a robust breakdown of various factors affecting a gate-based algorithm, including state preparation, oracle, connectivity, circuit rewriting, transpiling, and readout, and designed a calibration matrix to quantitatively evaluate the quantum error.

While currently available quantum devices are unable to perform quantum arithmetic beyond very small scales, the study of quantum noise in NISQ hardware has allowed rapid advancement in the development of tunable noise models. These models offer the unique opportunity for insight into the performance of quantum arithmetic algorithms by allowing the impact of different sources of noise to be isolated. In a preliminary work, we simulated the dependency of unsigned quantum Fourier Addition (QFA) and Quantum Fourier Multiplication (QFM) on one- and two-qubit (1q- and 2q-) gate error rates, on the approximation depth in the QFT, and on the number of superposed integer states involved in the operands (which we refer to as a quantum integer's order of superposition) \cite{basili2021performance}. We used noise models designed to mirror the performance of IBM superconducting quantum computers, and included perfect, noise-free simulations as a point of comparison to differentiate the noise introduced by the approximate QFT. 

Since the publishing of the preliminary results, we have performed the same calculations but for signed QFA (sQFA), and completely regenerated the previous QFA results after discovering minor bugs in the original results. Thus, the present work represents the culmination of the preliminary work's verified conclusion, with the inclusion of sQFA for comparison to the unsigned results. The isolation of other sources of error, such as thermal relaxation and qubit measurement, their simultaneous simulation with 1q-/2q- gate errors, as well as the impact of error mitigation and extrapolation from integers to real numbers, remain deferred to a future work. The consideration of more system-efficient integer-encoding schemes conducive to quantum arithmetic remain similarly deferred.

Finally, though not discussed, modular and non-modular forms of QFT-based quantum integer subtraction, division, exponentiation, and extensions to vector, matrix, and other tensor operations, rely on slight alterations to the same underlying algorithm, and, consequently, many of the considerations discussed here apply to them as well \cite{draper2000addition,2008varez,ruiz2017quantum,babu2017cost,2020sahin}. Our results also provide insight regarding the QFT, AQFT, and quantum algorithms that rely upon them.

This paper is organized as follows. Sec.~\ref{sec_related} provides a brief summary of contemporary advancements related to quantum arithmetic using both QFT and non-QFT based approaches. Sec.~\ref{sec_QFT} briefly describes our implementation of the AQFT. In Sec.~\ref{sec_QArith} we illustrate Quantum Fourier Addition and Quantum Fourier Multiplication, and describe our implementation. In Sec.~\ref{sec_results}, we provide specific details on the generation, collection, and presentation of performance data considered in this study, and discuss their interpretation. 
Finally Sec.~\ref{sec_summary} summarizes this work and discusses its future development.

\section{Contemporary Works in Quantum Arithmetic}
\label{sec_related}

We review and summarize recent related works regarding quantum arithmetic, including both Quantum Fourier Transform (QFT) and non-QFT approaches, considering performance improvements and resource efficiency. Unlike those references regarding performance evaluation mentioned in the introduction, the works cited here focus on algorithmic improvement. We first discuss QFT approaches before moving on to non-QFT methods, including topics such as alternative paradigms, computational models, optimization, and innovations on specific arithmetic operators.

\citet{kurt2023qft} describe practical implementations and their scalability, offering a primitive QFT-based quantum arithmetic logic unit (qALU) for addition and NAND on IBM quantum hardware. In a following work they further proposed scalable QFT circuit designs capable of processing multiple inputs, investigating the potential benefits of using qudits (specifically ququarts) over qubits \cite{kurt2024scalable}. Ref.~\cite{yuan2023improved} developed an improved quantum comparator, applicable to both quantum-classical and quantum-quantum inputs, building upon conventional approaches to QFT-based addition, and extend this technique to modular arithmetic operations requiring only one ancilla qubit.

Quantum multiplication algorithms have received particular attention in recent years. Ref.~\cite{crimmins2023efficient} explored techniques for efficient integer multiplication utilizing the QFT. Ref.~\cite{ramezani2023quantum} introduced an approach relying on the convolution theorem, implemented via QFT and amplitude amplification, offering an advantageous time complexity compared to leading classical algorithms for specific cases of multiplication.

Beyond QFT-centric approaches, notable research exists investigating alternative quantum arithmetic paradigms and optimization techniques. Comprehensive reviews, such as that provided in Ref.~\cite{wang2024comprehensive}, survey the diverse landscape of quantum arithmetic circuits, describing several strategies for fundamental operations. Ref.~\cite{chen2024quantum} proposed an alternative computational model for constructing quantum arithmetic operations based on Quantum Signal Processing. Ref.~\cite{joshi2023quantum} applied optimization frameworks like ZX calculus to reduce resource requirements for arithmetic circuits including multipliers.

A considerable amount of recent effort has focused on optimizing quantum operations using non-QFT methods. Ref.~\cite{wang2024optimal} achieved optimal logarithmic Toffoli-depth for quantum adders by systematically exploring classical carry-propagation structures adapted for quantum circuits, and Ref.~\cite{babu2022higher} developed higher-radix carry-lookahead adder designs. Ref.~\cite{gidney2018halving} introduced a notable optimization using temporary logical-AND gates, effectively halving the T-gate cost for standard quantum addition compared to prior non-QFT constructions. Ref.~\cite{wang2024boosting} considered boosting quantum slow divider efficiency through systematic design-space exploration, providing insight on the impact of selecting efficient adder sub-blocks to more complex operations. 

Research into quantum multipliers and dividers through various non-QFT techniques also exists. Ref.~\cite{gidney2024fast} presented a fast, zero-ancilla multiplication algorithm achieving sub-quadratic gate complexity without relying on the QFT. Ref.~\cite{zhan2023quantum} proposed a novel multiplier using an exponent adder design that encodes operands into logarithmic qubit states, offering a different trade-off space. Refs.~\cite{cao2019designs} and \cite{yang2022circuit} provided circuit-level optimizations targeting specific gate counts (T-gates, CNOTs) for quantum multipliers and dividers. Ref.~\cite{gouzien2020improving} specifically focused on methods for reducing T-gate counts and improving their distribution within standard multiplier circuits. Specialized techniques like windowed arithmetic for modular exponentiation, which offer space-time tradeoffs using classical memory lookups, have seen recent optimization efforts in Ref.~\cite{luongo2025optimized}, improving costs relevant to quantum cryptographic analysis.

Finally, notable efforts have been made to extend quantum arithmetic to floating-point numbers and adapt designs for practical considerations with near-term hardware. Ref.~\cite{ao2023implementation} details implementations of quantum floating-point adders based on standard integer arithmetic blocks. Ref.~\cite{haner2018quantum} explored both automated synthesis from classical hardware descriptions (Verilog) and manual optimization approaches for constructing floating-point circuits. Recognizing the connectivity constraints of current quantum devices, Ref.~\cite{li2024feasible} proposed feasible QALU designs for elementary operations restricted to nearest-neighbor interactions, making them more suitable for 2D qubit layouts typical of near-term hardware.
\section{Phase-encoding and the Quantum Fourier Transform}
\label{sec_QFT}

We begin by specifying the quantum integers (q-integers) discussed in this work as a mathematical construct, and how we encode them as a quantum data structure. We consider the space of integer states $\ket{i}$ with $i\in \mathbb{Z}$. A q-integer $y$ is thus defined as a superposition of integer states, 
$$\ket{y}=\overset{k}{\underset{i}{\sum}}p_{i}\ket{i},$$ 
where 
$$\overset{k}{\underset{i}{\sum}}p_{i}^{2} = 1$$ 
is the sum of all integer probabilities, $p_i^2$, and must add to unity. We refer to a superposition of $j$ unique integers with $j$ respective nonzero probability amplitudes as an order-$j$ q-integer.

The representation of q-integers can be achieved by several means, and their optimal implementation depends on the required computational model \cite{fillinger2013data}. 
For our purposes, we constrain the current discussion to dynamic q-integers - that is, q-integers whose values can be updated by methods without collapsing superposition - realized on quantum registers of $n$ qubits. We encode q-integers in two's complement, though the results of this work may be extended to fixed-point binary representations of numbers in general. The maximum order of superposition for an $n$-qubit q-integer is therefore limited by the size of the register's computational basis, $N=2^n$, or an order-$N$ q-integer.

We facilitate our discussion using the quantum circuit paradigm as our starting point. 
Similar to how one may encode integers classically in the computational basis as bitstrings, quantum superposition enables one to encode numbers in the relative phases of qubits. Since linear phase shifts are a fundamental operation of most quantum computers, this encoding is particularly convenient for performing basic arithmetic operations. 

We refer to this basis of relative phases as the Quantum Fourier Basis (QFB). The QFT is an $\mathcal{O}(n\log_2n)$ quantum algorithm that performs a Fourier Transform of quantum mechanical amplitudes, and serves to transform numbers encoded in the computational basis to the QFB.
Whereas the discrete Fourier Transform over classical fields involves an expansion that mixes vectors of complex numbers, the QFT involves the expansion of the computational basis $\ket{y}=\ket{0},\ \ket{1}, \ldots ,\ \ket{N-1}$ into mixtures of QFB states $\ket{k}=\ket{\phi(0)},\ \ket{\phi(1)}, \ldots ,\ \ket{\phi(N-1)}$ (or vice-versa), as
\begin{equation}
   \mathcal{QFT} \ket{y} = \frac{1}{\sqrt{N}}\sum_{k=0}^{N-1} e^{\imag 2\pi yk/N} \ket{k}
\end{equation} 
for initial integer state $\ket{y}$ encoded in the computational basis with $n$ qubits and $N=2^n$ unique states. The corresponding inverse QFT is defined as
\begin{equation}
    \mathcal{QFT}^{-1} \ket{k} = \frac{1}{\sqrt{N}}\sum_{y=0}^{N-1} e^{-\imag 2\pi yk/N} \ket{y}.
\end{equation}

When we apply the QFT to the state $\ket{y} = \ket{y_1} \otimes \ket{y_2} \otimes \ldots \otimes \ket{y_n}$ (or as a binary string $y = ``y_n y_{n-1}  \ldots y_1"$ such that $y = y_1 2^0 + y_2 2^1 + \ldots + y_n 2^{n-1}$), one obtains the result:
\begin{equation}
\label{eq_QFT2}
\begin{aligned}
    \mathcal{QFT} \ket{y} = & \frac{1}{2^{n/2}}\bigotimes_{q=1}^{n}\left(\ket{0}+ e^{\imag 2\pi 
    \left(\sum_{k=1}^{q} y_k/2^{q-k+1}\right)
    } \ket{1}\right)_q 
     = & \frac{1}{2^{n/2}}\bigotimes_{q=1}^{n}\left(\ket{0}+ e^{\imag 2\pi 
    [0.y]_{q,1}
    } \ket{1}\right)_q,
\end{aligned}
\end{equation}
where the parenthetical subscript on the right-hand side denotes the $q$th qubit's state, and 
we rewrite the summation in the exponent using the binary-fraction shorthand 
$[0.y]_{i,j}=0.y_i y_{i-1} \ldots y_{j}= y_i/2+y_{i-1}/4 + \cdots + y_j/2^{i-j+1}$ for convenience. 

If we consider from this the action of the QFT on each qubit, one notices that, after applying the QFT, the first qubit (which originally had the state $\ket{y_1}$) will possess a state where the phase summation on the right-hand side of Eq.~\eqref{eq_QFT2} reduces to only one term, $[0.y]_{1,1}=y_1/2$. Notably, this means the first qubit's final state only depends on its own initial state. Meanwhile, the $y_2$ qubit will have two terms, $[0.y]_{2,1}=y_2/2+y_1/4$, for its sum of phases, and thus depends on the initial states of both itself and the first qubit. Finally, the $n$th qubit's final state will involve a sum of n terms, and depends on the initial states of all n qubits. If we consider each term of these summations as a single phase rotation being controlled by a single qubit, the first qubit requires one such rotation, the second requires two, and so on. Further, the first rotation on a given qubit is controlled by itself, while each subsequent rotation is controlled by each qubit that comes before it.

To implement the QFT described by Eq.~\eqref{eq_QFT2} as a quantum circuit, we must construct the process just described in terms of unitaries. We therefore introduce the two-qubit controlled rotation phase gate $R_{l}$ and one-qubit Hadamard gate $H$ with which we shall construct the circuit. Using the basis ordering common to most textbooks for our matrix representation of quantum operators (sometimes called big-endian convention), we may write the unitary transformations
\begin{equation}
    R_{l} =\begin{bmatrix}
    1 & 0 & 0 & 0 \\
    0 & 1 & 0 & 0 \\
    0 & 0 & 1 & 0 \\
    0 & 0 & 0 & e^{\frac{\imag 2\pi }{2^l}}
    \end{bmatrix},
    H =\frac{1}{\sqrt{2}}\begin{bmatrix}
    1 & 1 \\
    1 & -1
    \end{bmatrix},
\end{equation}
where it is understood the dimensions span the corresponding Hilbert space specific to the given qubit(s) that the operator acts upon. Notably, both of these gates can be efficiently constructed with only one or two gates of the universal gate set used by current IBM superconducting quantum devices \cite{cross2018ibm}.

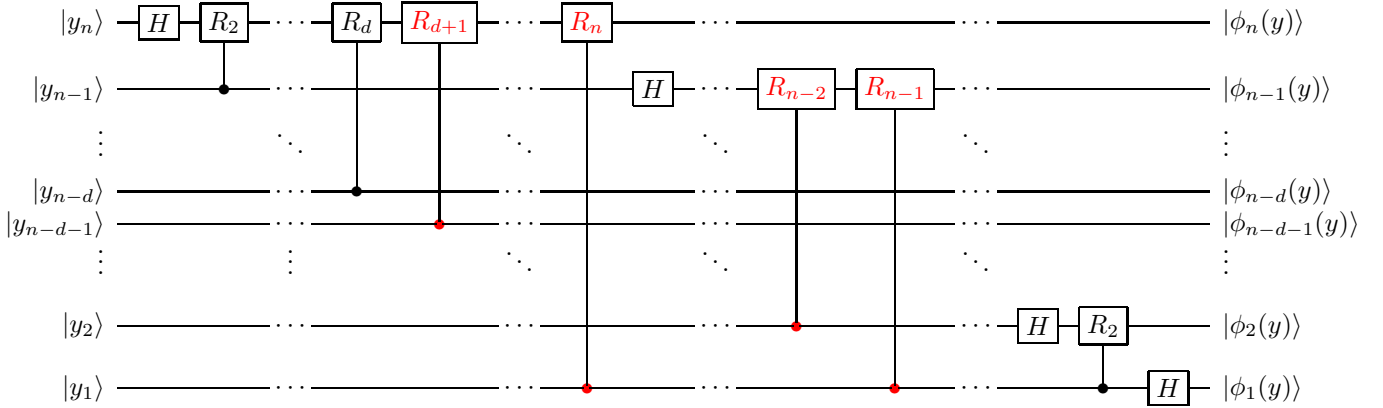
\begin{figure*}[t]
\centering
\[
\begin{array}{c}
\Qcircuit @C=0.8em @R=1.0em {
  \lstick{\ket{y_{n}}}   & \gate{H} & \gate{R_{2}} & \qw   & \cdots &    &   
  \gate{R_{d}}  &   \gatered{\color{red}R_{d+1}} & \qw     & \cdots &    &
  \gatered{\color{red}R_{n}}   &   \qw    & \qw &   \cdots &   &
  \qw                          & \qw                          &
  \qw & \cdots  &  & \qw      & \qw          & \qw     &
  \rstick{\ket{\phi_n(y)} }  \qw        \\
  \lstick{\ket{y_{n-1}}}  & \qw     & \ctrl{-1}    & \qw   & \cdots &    &   
  \qw             &    \qw                     & \qw     & \cdots &      & 
  \qw                          & \gate{H} & \qw &   \cdots &   & 
  \gatered{\color{red}R_{n-2}} & \gatered{\color{red}R_{n-1}} &
  \qw & \cdots  &  & \qw      & \qw          & \qw     &
  \rstick{\ket{\phi_{n-1}(y)}}   \qw       \\
  \lstick{\vdots }        &         &              &       & \ddots &    &   
                  &                            &         & \ddots  &     &
                               &          &     &   \ddots &   &
                               &                              &
      & \ddots  &  &          &              &         &
  \rstick{\vdots}\\
  \\
  \lstick{\ket{y_{n-d}}}  &   \qw   &   \qw        &   \qw & \cdots &    &
  \ctrl{-4}       &  \qw                       & \qw     & \cdots  &      &  
  \qw                          & \qw      & \qw &  \cdots &   & 
  \qw                          & \qw                         &
  \qw & \cdots  &  & \qw      & \qw          & \qw     &
  \rstick{\ket{\phi_{n-d}(y)} }  \qw        \\
  \lstick{\ket{y_{n-d-1}}}  &   \qw   &   \qw      &   \qw & \cdots &    &
  \qw             &  \ctrlred{-5}               & \qw     & \cdots  &     &  
  \qw                          & \qw      & \qw &  \cdots &   &
  \qw                          & \qw                          &
  \qw & \cdots  &  & \qw      & \qw          & \qw     &
  \rstick{\ket{\phi_{n-d-1}(y)} }  \qw        \\
  \lstick{\vdots }        &         &              &       & \vdots &    &   
                  &                            &         & \ddots  &     &
                               &          &     &   \ddots &   &
                               &                              &
      & \ddots  &  &          &              &         &
  \rstick{\vdots}\\
  \\
  \lstick{\ket{y_{2}}}    &   \qw   &   \qw        &   \qw & \cdots &    &
  \qw             & \qw                        & \qw     &  \cdots &     &
  \qw                          & \qw      & \qw &   \cdots &   &
  \ctrlred{-7}                   & \qw                          &
  \qw & \cdots  &  & \gate{H} & \gate{R_2}   & \qw     &  
  \rstick{\ket{\phi_2(y)} } \qw  \\
  \lstick{\ket{y_{1}}}    &   \qw   &   \qw        &   \qw & \cdots &    &
  \qw             & \qw                        & \qw     &  \cdots &     &
  \ctrlred{-9}                 & \qw      & \qw &   \cdots &   & 
    \qw                          & \ctrlred{-8}                  &
  \qw & \cdots  &  & \qw      & \ctrl{-1}   & \gate{H} &   
  \rstick{\ket{\phi_1(y)} } \qw
}
\end{array}
\]
  \caption{Generalized quantum circuit for the QFT. Gates removed when performing the AQFT at approximation depth $d$ are drawn in red.}
  \label{fig_QFT_circuit}
  \end{figure*}

To perform the QFT on a single $n$-qubit q-integer, we first apply the Hadamard gate to each qubit, followed by consecutive controlled-$R_{l}$ gates as shown in Fig.\ \ref{fig_QFT_circuit}. Repeating the procedure for all the qubits in the q-integer's register and including the normalization factor of $1/\sqrt{2}$, we obtain the qubit state shown on the right in Eq. \eqref{eq_QFT2}.


When a large number of qubits are involved, the conditional rotation $R_{l \rightarrow \infty}\rightarrow I$ of the QFT approaches the identity. On one hand, encoding in the QFB therefore requires that the accuracy of the phase shift performed by the physical quantum computer increases exponentially with $n$ to ensure addition via phase shifts remains truly linear at all scales. On the other hand, one rarely requires such accuracy for phase rotations between bits of sufficiently different binary order in most practical computations. This enables approximation that is particularly useful when working with noisy quantum devices, which is a primary motivation of the AQFT.

In the AQFT, we define a depth, $d$, which serves as an upper limit for the number of conditional rotation gates  applied to each qubit in the QFT circuit. We illustrate this in Fig.\ \ref{fig_QFT_circuit}, where gates present in the QFT but removed in the AQFT are marked in red. In comparison to Eq.~\eqref{eq_QFT2}, the AQFT takes the form 
\begin{equation}
\label{eq_AQFT}
    \mathcal{AQFT} \ket{y} = \frac{1}{2^{n/2}}\bigotimes_{q=1}^{n}\left(\ket{0}+ e^{\imag 2\pi 
    [0.y]_{q,d}
    } \ket{1}\right)_q.
\end{equation}

While the QFT requires $n(n+1)/2$ operations and possesses a gate complexity of $O(n^2)$, the AQFT at depth d reduces this to $(2n-d)(d-1)/2$. As derived by Barenco et al. \cite{1996barenco}, in the presence of decoherence and sufficiently large n, one expects the optimal depth of the AQFT to approximately approach $d\rightarrow \log_2 n$, requiring $O(n\log_2 n)$ operations. While the results in Section\ \ref{sec_results} support this, we do observe instances where, in certain performance regimes, the optimal depth can deviate from this expectation. 


\begin{figure*}[t]
\centering
\[
\begin{array}{c}
\Qcircuit @C=0.8em @R=1.0em {
      \lstick{\ket{x_{n}}}   & \ctrl{6}  &  \qw &   \qw & \cdots&    &   \qw &   \qw   &   \qw   &\ctrl{7} &   \qw   &   \qw &      \qw &  \qw    &   \qw & \qw &   \qw &   \qw   & \qw   &    \qw &   \qw &   \rstick{\ket{x_n}}  \qw        \\
  \lstick{\ket{x_{n-1}}}  &  \qw    &   \ctrl{5}  &    \qw  & \cdots&  &   \qw & \qw&  \qw    &    \qw   &\ctrl{6}    & \qw &   \qw   &  \qw  &       \qw  & \qw &    \qw   &\qw &\qw    & \qw  &   \qw  & \rstick{\ket{x_{n-1}}}\qw       \\
  \lstick{\vdots }  &   &   &   &   &   \ddots &   &    &   &   &   &   &  \ddots  &   &   &   &   &   &   &    & &  \rstick{\vdots}  \\
  \\
  \lstick{\ket{x_{2}}}    &   \qw &   \qw &   \qw &   \qw &   \qw &  \ctrl{2}     &  \qw &  \qw  &   \qw    &  \qw   &   \qw  &   \qw &   \qw   &  \ctrl{3}  &   \qw   & \qw  &    \ctrl{6} & \qw &  \qw   & \qw   &  \rstick{\ket{x_{2}}}\qw  
  \\
  \lstick{\ket{x_{1}}}  &   \qw &   \qw &   \qw &   \qw  &  \qw &  \qw &   \ctrl{1} &   \qw&   \qw   &   \qw &   \qw &   \qw  &   \qw    &   \qw &   \ctrl{2} &   \qw  &  \qw   & \ctrl{5}  &  \qw    &  \ctrl{6}   &    \rstick{\ket{x_{1}}} \qw
  \\
 \lstick{\ket{\phi_{n+1}(y)}}   &  \gate{R_2}    &   \gate{R_{3}}   &   \qw & \cdots&    &   \gate{R_{n}}   &   \gate{R_{n+1}}   &   \qw    &   \qw   &   \qw &   \qw &   \qw &  \qw    &   \qw & \qw & \qw &   \qw   & \qw   &    \qw &   \qw &   \rstick{\ket{\phi_{n+1}(x+y)}}  \qw      \\    
  \lstick{\ket{\phi_{n}(y)}}  &  \qw    &   \qw  &\qw  &  \qw &   \qw&  \qw    &    \qw &   \qw   & \gate{R_1} &   \gate{R_2} & \qw &  \cdots &   & \gate{R_{n-1}}  &   \gate{R_{n}}    &    \qw  & \qw &    \qw   &\qw &\qw  & \rstick{\ket{\phi_{n}(x+y)}}\qw       \\
  \lstick{\vdots }  &   &   &   &   &   &  &  &    &   &   &   &   \vdots  &   &   &   &   &   &   & & & \rstick{\vdots}  \\
  \\
  \lstick{\ket{\phi_2(y)}}    &   \qw &   \qw &   \qw &   \qw &   \qw &  \qw     &  \qw &  \qw  &   \qw    &  \qw   &   \qw  &  \qw  &   \qw &   \qw &   \qw   & \qw & \gate{R_1} & \gate{R_2}   & \qw   & \qw   &  \rstick{\ket{\phi_2(x+y)}}\qw  \\
  \lstick{\ket{\phi_1(y)}}  &   \qw &   \qw &   \qw &   \qw  &  \qw  &  \qw &   \qw &   \qw &   \qw &   \qw &   \qw &   \qw &   \qw   &   \qw &   \qw  &  \qw   & \qw  &   \qw    &   \qw &   \gate{R_1}    &  \rstick{\ket{\phi_1(x+y)}} \qw 
  }
  \end{array}
  \]
  \caption{The generalized addition step of the QFA circuit, being applied in the transform domain.}
  \label{fig_Qadder}
  \end{figure*}
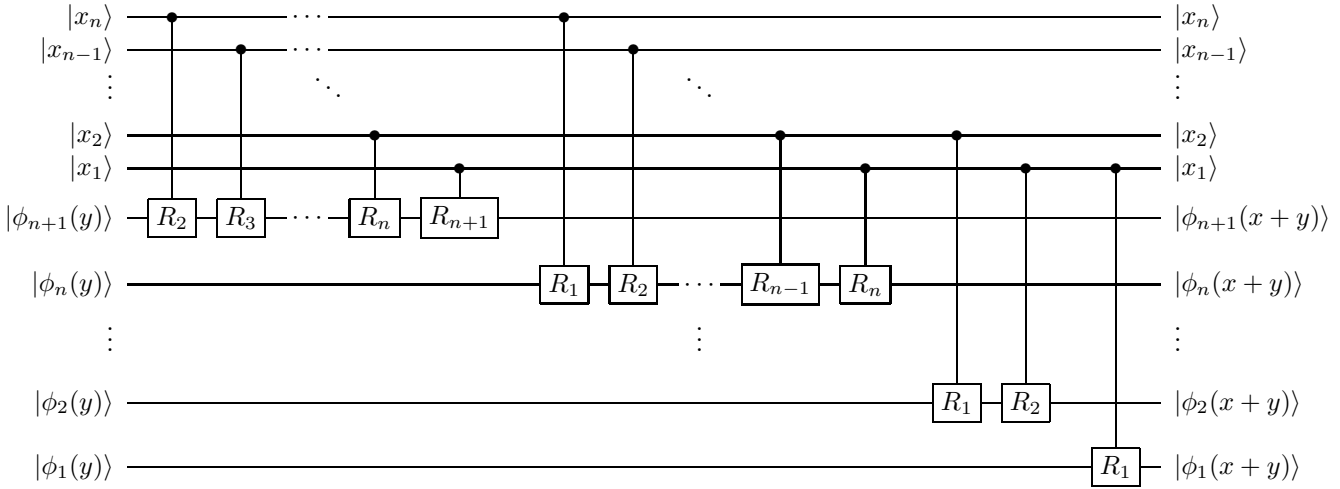

\section{Quantum arithmetic operations}
\label{sec_QArith}
We now consider two operand q-integers, $\ket{x}$ and $\ket{y}$ of $n$ and $m$ qubits, respectively. We discuss our implementation of Quantum Fourier Addition (QFA) first, as it is required for our implementation of Quantum Fourier Multiplication (QFM). 

The first step of QFA is to transform one of the addends (we will choose $y$) into the Fourier basis with the QFT, as discussed in Sec.~\ref{sec_QFT}. The `addition step' can then be performed through a sequence of conditional rotations that add the phase-encoded magnitude of $\ket{x}$ to $\ket{\phi(y)}$. The process is completed by applying the inverse QFT to $\ket{\phi(x+y)}$, to return that register to the computational basis.
Throughout this process, $\ket{x}$ remains in the computational basis. 

Whereas, for the QFT, the conditional rotations on a given qubit (i.e., $y_i$) are controlled entirely by the other qubits defining $\ket{y}$, the rotations performed in the addition step are controlled by the qubits of $\ket{x}$ following 
\begin{align}
    \mathcal{QFA}  \frac{1}{\sqrt{N}}\sum_{k=0}^{N-1} e^{\imag 2 \pi y k/N}\ket{x} \ket{k} 
     =  \frac{1}{\sqrt{N}}\sum_{k=0}^{N-1}e^{\imag 2 \pi x k/N} e^{\imag 2 \pi y k/N}\ket{x} \ket{k}. 
\end{align}
The quantum circuit for the QFA addition step is shown in Fig. \ref{fig_Qadder}. Just like with the QFT, the deeper rotations of the addition step approach the identity, and it has been speculated that a similar cutoff on depth in the add step may improve performance in the presence of noise \cite{draper2000addition}. However, unlike with the AQFT where the cutoff impacts only the linearity of the transform domain, a similar cutoff in the addition step directly impacts the phase shifts being added themselves. This, in addition to the fact that the same cutoff in the addition step will only remove half as many gates as the AQFT does in QFA, leads us to expect the benefits of such an approximation will be less apparent than the AQFT (particularly at lower $n$). Thus, we have opted to consider only the AQFT first, and leave the consideration of an approximate addition step for a future work.

\begin{figure*}[t]
\centering
\[
\begin{array}{c}
\Qcircuit @C=0.8em @R=1.0em {
  \lstick{\ket{x}} & \qw {/} & \ustick{n} \qw & \qw & \ctrl{2} & \rstick{\ket{x}} \qw & & & 
    \lstick{\ket{x}} & \qw {/} & \ustick{n} \qw & \qw & \qw & \ctrl{2} & \qw & \rstick{\ket{x}} \qw \\
        &    &  &     &   &                    & \push{\rule{2.9em}{0em}=\rule{1.9em}{0em}} & & & & & &\\
  \lstick{\ket{y}} & {/} \qw & \ustick{n+1} \qw & \qw & \gate{QFA} & \rstick{\ket{x+y}} \qw & & &
  \lstick{\ket{y}}   & \qw {/} & \ustick{n+1} \qw & \qw & \gate{QFT} & \gate{add} & \gate{QFT^{-1}} & \rstick{\ket{x+y}} \qw
}
\ \\
\  \\
\hline
\hline
\\
\Qcircuit @C=0.8em @R=1.0em {
  \lstick{\ket{x}} & \qw {/} & \ustick{n} \qw & \qw & \multigate{6}{sQFA} & \qw {/} & \ustick{n} \qw & \rstick{\ket{x}} \qw & & & 
    \lstick{\ket{x_n}} & \qw & \qw & \qw & \qw & \multigate{6}{add} & \qw & \rstick{\ket{x_n}} \qw \\
     & & & & & & & & & &
    \lstick{\vdots} & & & & & & & \rstick{\vdots} \\
     & & & & & & & & & &
    \lstick{\ket{x_1}} & \qw & \ctrl{3} & \ctrlo{3} & \qw & \ghost{add} & \qw & \rstick{\ket{x_1}} \qw \\
  \lstick{\ket{y}} & {/} \qw & \ustick{n} \qw & \qw & \ghost{sQFA} & \qw {/} & \ustick{n+1} \qw & \rstick{\ket{x+y}} \qw & \push{\rule{2.9em}{0em}=\rule{1.9em}{0em}} & &
  \lstick{\ket{y_n}} & \qw & \qw & \qw & \multigate{3}{QFT} & \ghost{add} & \multigate{3}{QFT^{-1}} & \rstick{\ket{(x+y)_{n+1}}} \qw \\
     & & & & & & & & & &
    \lstick{\vdots} & & & & & & & \rstick{\vdots} \\
     & & & & & & & & & &
    \lstick{\ket{y_1}} & \qw & \ctrlo{1} & \ctrl{1} & \ghost{QFT} & \ghost{add} & \ghost{QFT^{-1}} & \rstick{\ket{(x+y)_2}} \qw \\
  \lstick{\ket{0}} & \qw & \qw & \qw & \ghost{sQFA} & & & & & &
  \lstick{\ket{0}} & \qw & \targ & \targ & \ghost{QFT} & \ghost{add} & \ghost{QFT^{-1}} & \rstick{\ket{(x+y)_1}} \qw \\
}
\ \\
\  \\
\hline
\hline
\\
\Qcircuit @C=0.8em @R=1.0em {
  \lstick{\ket{c}} & \qw & \qw & \qw & \ctrl{1} & \rstick{\ket{c}} \qw & & & 
    \lstick{\ket{c}} & \qw & \qw  & \qw & \ctrl{2} & \ctrl{1} & \ctrl{2} & \rstick{\ket{c}} \qw \\
  \lstick{\ket{x}} & \qw {/} & \ustick{n} \qw & \qw & \ctrl{1} & \rstick{\ket{x}} \qw & \push{\rule{2.9em}{0em}=\rule{1.9em}{0em}} & & 
    \lstick{\ket{x}} & \qw {/} & \ustick{n} \qw & \qw & \qw & \ctrl{1} & \qw & \rstick{\ket{x}} \qw \\
  \lstick{\ket{y}} & {/} \qw & \ustick{n+1} \qw & \qw & \gate{cQFA} & \rstick{\ket{x+y}} \qw & & &
  \lstick{\ket{y}}   & \qw {/} & \ustick{n+1} \qw & \qw & \gate{cQFT} & \gate{cadd} & \gate{cQFT^{-1}} & \rstick{\ket{x+y}} \qw
}
  \end{array}
\]
  \caption{The QFA (top), sQFA (middle), and cQFA (bottom) composite gates.}
  \label{fig_qadd_gate1}
\end{figure*}

Once the addition step is complete and 
$\ket{\phi(y)}\rightarrow \ket{\phi(x+y)}$, 
one recovers the sum by transforming the output back to the computational basis using the inverse QFT:
\begin{equation}
    \begin{aligned}
    \mathcal{QFT}^{-1} \frac{1}{\sqrt{N}}\sum_{k=0}^{N-1}e^{\imag 2 \pi x k/N} e^{\imag 2 \pi y k/N}\ket{x} \ket{k} 
    & = \frac{1}{N} \sum_{k,l=0}^{N-1} e^{\imag 2\pi (x+y)k /N}e^{ -\imag 2\pi kl/N} \ket{x}\ket{l} \\
    & = \ket{x} \ket{(x+y) \mod N}.
    \end{aligned}
\end{equation}
Applying the QFT, addition step, and inverse QFT together, we define the full QFA composite gate in the top of Fig.\ \ref{fig_qadd_gate1}.
To perform non-modular addition, one simply ensures the register being updated ($\ket{y}$ in our case) is large enough to avoid overflow by including one or more ancillary qubits (i.e. $m=n+1$. For $m > n+1$, the QFA process only involves the first $n+1$ qubits of $y$). 

In unsigned QFA, the aforementioned ancillary qubit appended to $y$ simply becomes the new most significant qubit of the new q-integer $(x+y)$. Meanwhile, in signed QFA (sQFA) encoded in two's compliment, the ancillary qubit becomes the new sign qubit, requiring two additional Toffoli gates in the sQFA compared to the QFA, as shown in the middle diagram of Fig. \ref{fig_qadd_gate1}. The added gate logic can best be understood by considering $x$ and $y$ when in single integer states: if the sign of $x$ and $y$ match, neither gate flips the ancillary qubit. If they do not match (i.e. one is positive and one is negative), exactly one of the two Toffoli gates flips the ancillary, effectively causing the \textit{add} process to result in the difference of the q-integer phases instead of their sum. The same logic applies to all potential superposed q-integer states as well. Strategies for modular addition of signed and unsigned integers with the QFT have also been developed \cite{ruiz2017quantum,2020sahin}. 

A few different approaches for quantum multiplication with the QFT have been proposed \cite{2008varez,2020sahin}. For the purposes of this study, we implement a QFM approach similar to the weighted-sum strategy for unsigned integers presented by Ruiz-Perez \cite{ruiz2017quantum}.
Unlike QFA described above, in which one of the addends is updated, QFM preserves both multiplicand states, and instead updates a separate product register (which is generally initialized to zero). To avoid any potential overflow, the product register must be at least as large as the combined sizes of the two multiplicand registers. The process involves applying the QFA iteratively on a varying subset of the product register, while being controlled by the multiplicand's qubits.

In order to define the QFM circuit explicitly, we must first introduce the controlled-controlled-phase rotation gate, $cR_l$, and controlled-Hadamard gate, $cH$, which we will use to define the controlled QFT (cQFT) and controlled-addition step (cadd) required by the QFM.

\begin{equation}
    cR_{l} =\begin{bsmallmatrix}
    1 & 0 & 0 & 0 & 0 & 0 & 0 & 0 \\
    0 & 1 & 0 & 0 & 0 & 0 & 0 & 0 \\
    0 & 0 & 1 & 0 & 0 & 0 & 0 & 0 \\
    0 & 0 & 0 & 1 & 0 & 0 & 0 & 0 \\
    0 & 0 & 0 & 0 & 1 & 0 & 0 & 0 \\
    0 & 0 & 0 & 0 & 0 & 1 & 0 & 0 \\
    0 & 0 & 0 & 0 & 0 & 0 & 1 & 0 \\
    0 & 0 & 0 & 0 & 0 & 0 & 0 & exp\left({\frac{\imag 2\pi }{2^l}}\right)
    \end{bsmallmatrix},
    cH =\begin{bsmallmatrix} \scriptstyle
    1 & 0 & 0 & 0 \\
    0 & 1 & 0 & 0 \\
    0 & 0 & \frac{1}{\sqrt{2}} & \frac{1}{\sqrt{2}} \\
    0 & 0 & \frac{1}{\sqrt{2}} & -\frac{1}{\sqrt{2}}
    \end{bsmallmatrix}
\end{equation}

Replacing the $R_l$ and $H$ gates with $cR_l$ and $cH$ in Figs.~\ref{fig_QFT_circuit}~\&~\ref{fig_Qadder}, we define the cQFT, cadd, and cQFT$^{-1}$ circuits just as before, but now with one additional control qubit, $c$, controlling every gate of the process. We define the controlled QFA (cQFA) composite gate in the bottom of Fig.\ \ref{fig_qadd_gate1}. 

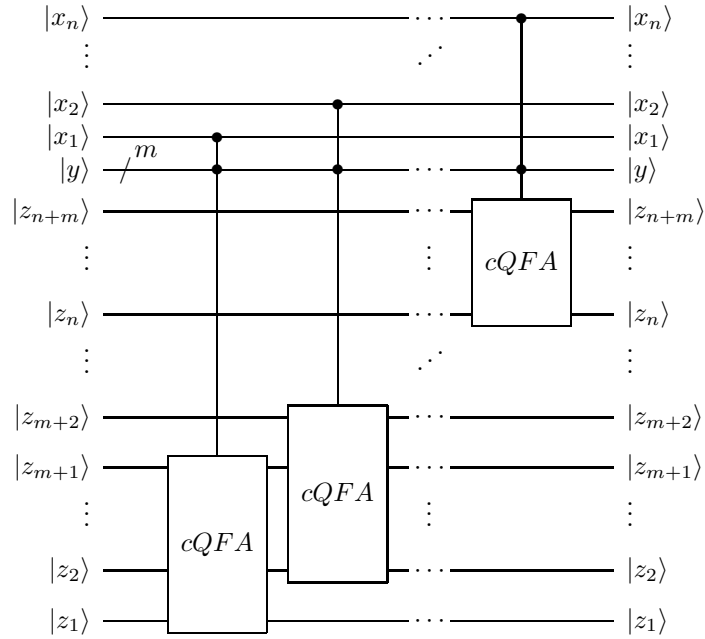
\begin{figure*}[t]
\[
\begin{array}{c} 
\Qcircuit @C=0.8em @R=1.0em {
      \lstick{\ket{x_{n}}} & \qw  &  \qw & \qw &   \qw & \qw & \cdots&    &   \ctrl{5}   &   \qw &   \rstick{\ket{x_n}}  \qw        \\
  \lstick{\vdots }  &   &   &   &  & & \vdots &   &    &   &  \rstick{\vdots}  \\
  \\
  \lstick{\ket{x_{2}}}  &   \qw & \qw &   \qw      & \ctrl{2} & \qw & \cdots &     &  \qw &  \qw &  \rstick{\ket{x_{2}}}\qw  
  \\
  \lstick{\ket{x_{1}}}  &   \qw &   \qw & \ctrl{1} &   \qw  & \qw & \cdots &   &   \qw &   \qw &    \rstick{\ket{x_{1}}} \qw
  \\
   \lstick{\ket{y}}  &    {/} \qw & \ustick{m} \qw & \ctrl{8} & \ctrl{7} &\qw &  \cdots &  &   \ctrl{1} & \qw   &    \rstick{\ket{y}} \qw
  \\
 \lstick{\ket{z_{n+m}}}   &  \qw    &   \qw   &   \qw & \qw & \qw &  \cdots&    & \multigate{3}{cQFA} &   \qw &   \rstick{\ket{z_{n+m}}}  \qw      \\
\lstick{\vdots }  &   &   &   &   &   & \vdots & & &    & \rstick{\vdots}  \\
  \\
 \lstick{\ket{z_{n}}}   &  \qw    &   \qw   &   \qw & \qw  & \qw & \cdots&    &  \ghost{cQFA}   &   \qw &   \rstick{\ket{z_{n}}}  \qw      \\
  \lstick{\vdots}  &   &   &  &   &   & \vdots & & &    & \rstick{\vdots}  \\
  \\
 \lstick{\ket{z_{m+2}}}   & \qw     &   \qw   &   \qw & \multigate{4}{cQFA} & \qw & \cdots&   &   \qw & \qw &  \rstick{\ket{z_{m+2}}}  \qw      \\    
  \lstick{\ket{z_{m+1}}}  &  \qw    &   \qw  & \multigate{4}{cQFA} & \ghost{cQFA} & \qw &   \cdots & &    \qw &   \qw  & \rstick{\ket{z_{m+1}}}\qw       \\
  \lstick{\vdots }  &   &   &  &  &   & \vdots & & &    & \rstick{\vdots}  \\
  \\
  \lstick{\ket{z_2}}    &   \qw &   \qw & \ghost{cQFA} & \ghost{cQFA} &   \qw & \cdots &     &  \qw &  \qw   &  \rstick{\ket{z_2}}\qw  \\
  \lstick{\ket{z_1}}  &   \qw &   \qw & \ghost{cQFA} & \qw &  \qw  &  \cdots  & &   \qw &   \qw    &  \rstick{\ket{z_1}} \qw 
  }
  \end{array}
\]
  \caption{The generalized QFM circuit, drawn such that $m<n-1$. If $n=m$, then $\ket{z_n} = \ket{z_{m+1}}$ and the diagram shifts, and further shifts if $m>n$.}
  \label{fig_Qmult}
  \end{figure*}

We express the QFM circuit in terms cQFA gates in Fig.~\ref{fig_Qmult}. We denote the product register of $n+m$ qubits as $\ket{z}$. 
Here, the qubits of $\ket{x}$ take the role of the additional control qubit. 

The QFM process is perhaps easiest to understand when considering a classical case, where the q-integers are each in a single integer state, and QFM resembles classical binary multiply. In that case, the $i$th cQFA step adds $x_i2^{i-1}y$ to the product register, where $x_i$ is 0 or 1, and $y$ is a single integer.


\section{Results and Discussion} \label{sec_results}

\begin{figure*}[h!]
\centering
\includegraphics[width=1.0\textwidth]{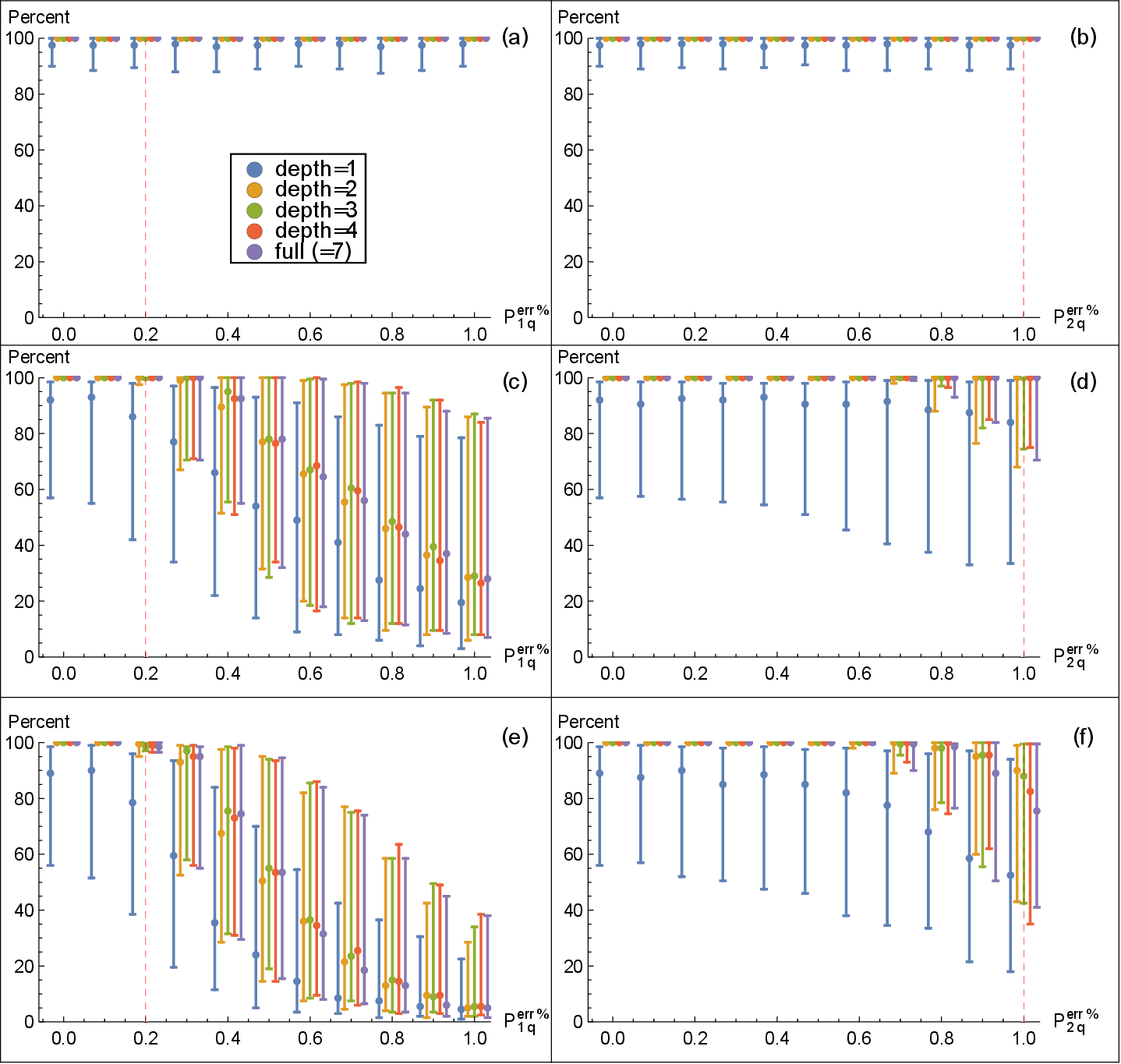}

  \caption{Success rates (vertical axes) of \textbf{signed} quantum addition of two 8-qubit q-integers performed in the Quantum Fourier Basis with varying gate error rates, number of superposed addend state, and AQFT approximation depth; see Sec.~\ref{sec_results} for details. The left/right column presents results with varying 1q-gate/2q-gate error rate, ($P_{1q}^{err\%}/P_{2q}^{err\%}$) on the horizontal axes, respectively. The top/middle/bottom row shows results for (1:1)/(1:2)/(2:2) superposed addend states. Color-coded results for each depth of AQFM approximation depth are clustered along the horizontal axes and are only shifted apart horizontally for visibility; the alignment of the center-most point of a cluster reflects the gate error rate used for all calculations in that cluster. The `full' approximation depth designates that the full QFT circuit is performed in those simulations.}
  \label{fig_sqfe}
  \end{figure*}

\begin{figure*}[h!]
\centering
\includegraphics[width=1.0\textwidth]{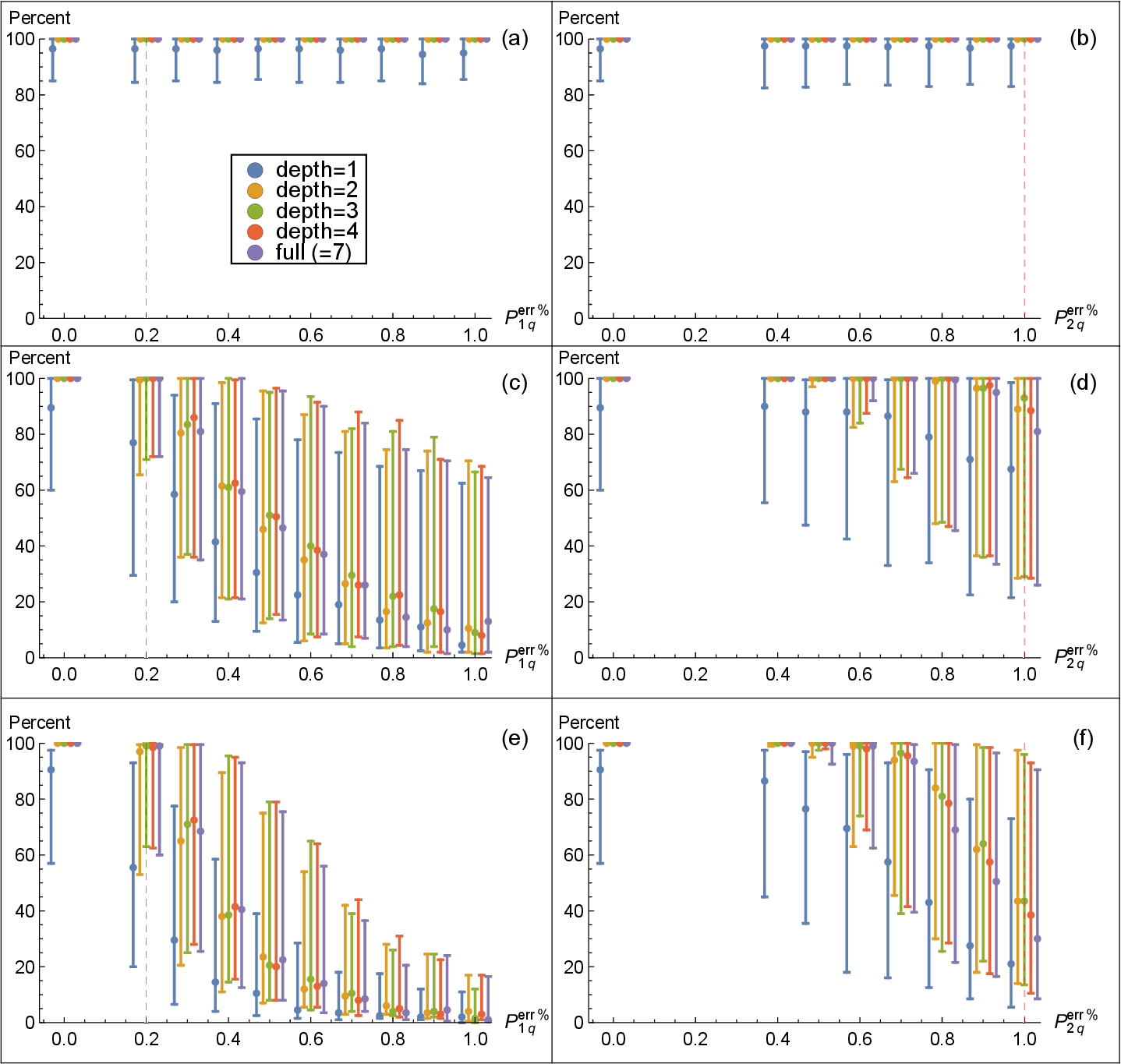}

  \caption{Success rates (vertical axes) of \textbf{unsigned} quantum addition of two 8-qubit q-integers performed in the Quantum Fourier Basis with varying gate error rates, number of superposed addend state, and AQFT approximation depth; see Sec.~\ref{sec_results} for details. The left/right column presents results with varying 1q-gate/2q-gate error rate, ($P_{1q}^{err\%}/P_{2q}^{err\%}$) on the horizontal axes, respectively. The top/middle/bottom row shows results for (1:1)/(1:2)/(2:2) superposed addend states. Color-coded results for each depth of AQFM approximation depth are clustered along the horizontal axes and are only shifted apart horizontally for visibility; the alignment of the center-most point of a cluster reflects the gate error rate used for all calculations in that cluster. The `full' approximation depth designates that the full QFT circuit is performed in those simulations. This figure represents data generated after minor bugs were discovered in the generation of similar data presented in \cite{basili2021performance}. Despite the corrections and being a new data set, the new results are nearly indistinguishable from the former, suggesting the bugs only affected edge cases.}
  \label{fig_qfe}
  \end{figure*}

\begin{figure*}[h!]
\centering
\includegraphics[width=1.0\textwidth]{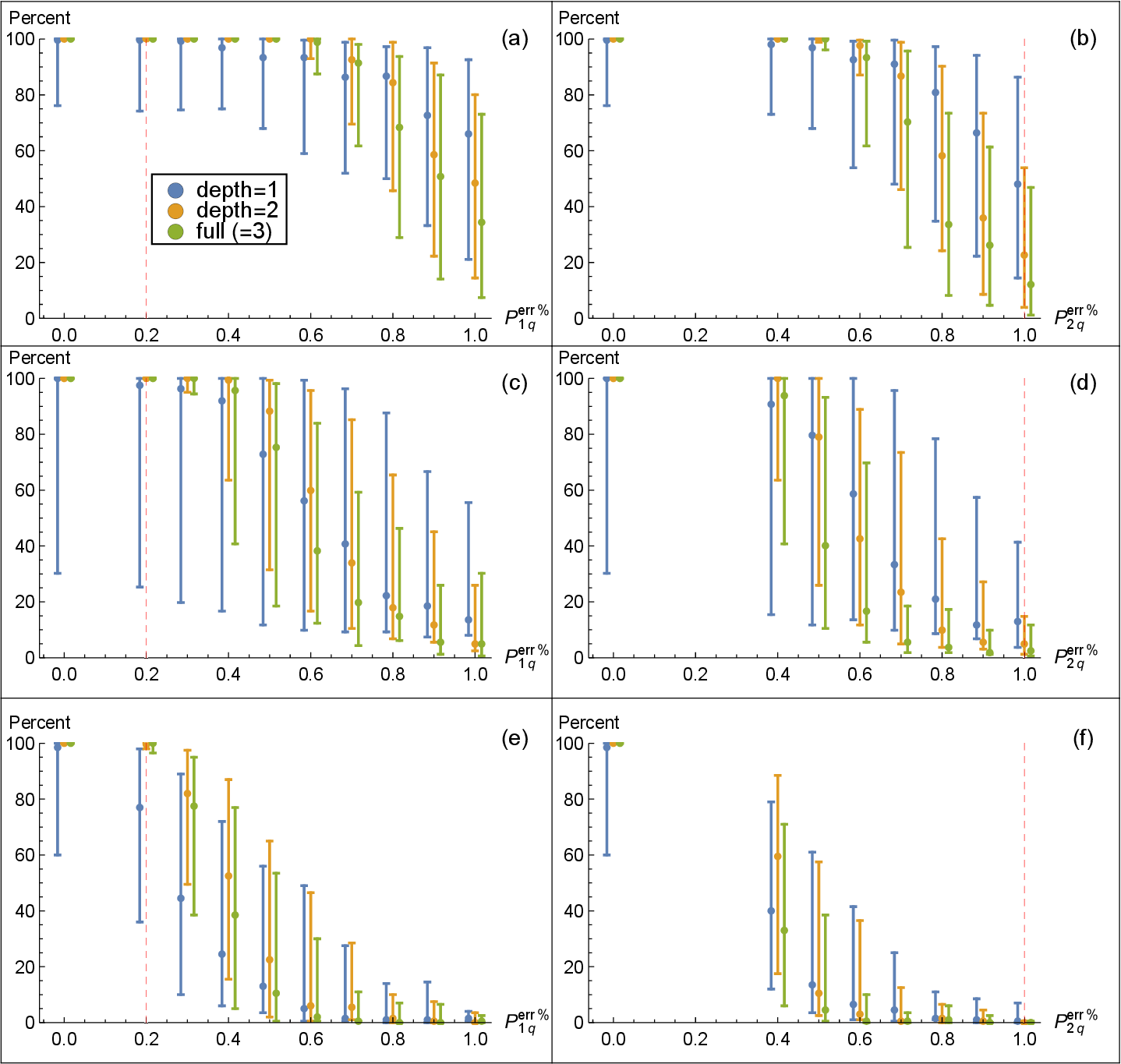}

  \caption{Success rates (vertical axes) of \textbf{unsigned} quantum multiplication of two 4-qubit q-integers performed in the Quantum Fourier Basis with varying gate error rates, number of superposed multiplicand state, and AQFT approximation depth; see Sec.~\ref{sec_results} for full details. The left/right column presents results with varying 1q-gate/2q-gate error rate ($P_{1q}^{err\%}/P_{2q}^{err\%}$) on the horizontal axes, respectively. The top/middle/bottom row shows results for (1:1)/(1:2)/(2:2) superposed multiplicand states. Color-coded results for each depth of AQFM approximation depth are clustered along the horizontal axes and are only shifted apart horizontally for visibility; the alignment of the center-most point of a cluster reflects the gate error rate used for all calculations in that cluster. The `full' approximation depth designates that the full QFT circuit is performed in those simulations. Figure reproduced from the results of \cite{basili2021performance}.}
  \label{fig_qfm}
  \end{figure*}

We present performance results of simulated signed and unsigned quantum Fourier addition and multiplication for varying approximation depths, 1q-/2q-gate error rates, and operand's order of superposition. Fig. \ref{fig_sqfe} presents results from signed addition between two $n=8$ q-integers, Fig. \ref{fig_qfe} presents results from unsigned addition between two $n=8$ q-integers, and Fig. \ref{fig_qfm} presents the results of multiplication between two $n=4$ q-integers. 

We use the same organizational structure in all three figures. The left column presents results with varying 1q-gate error rate, while the right column shows results with varying 2q-gate error rate. The top row shows results for operations where both operands are order-1 q-integers (henceforth referred to as a 1:1 operation). The middle row reflects results where one operand is order-1 and the other is order-2 (a 1:2 operation; in the case of addition, the order-2 addend is always stored on the qubit register that is being updated). The bottom row reflects results where both operands are order-2 q-integers (2:2 operation). For all superposed integer states, the probability amplitude is evenly distributed between each state at initialization. We also use the same randomly-generated set of operand states for calculating results of both varying 1q-gate error and varying 2q-gate error.

Results in the figures are clustered along the horizontal axis based on their gate error rate. All points within the same cluster correspond to the same error rate, and are only shifted apart horizontally for visibility. The alignment of the center-most point of a cluster on the x-axis reflects the gate error rate used for all calculations in that cluster (for instance, all points in the right-most cluster correspond to results simulated with a 1.0\% gate error rate, then 0.9\%, for the next cluster, and so on). Results are color-coded based on the approximation depth used in those calculations and correspond to the legend in panel (a) of the respective figure. `Full' designates that the full QFT and inverse QFT circuits are performed in those simulations.

On the x-axis for each figure, we vary either 1q-gate or 2q-gate error rates (denoted $P^{err}_{1q}$ and $P^{err}_{2q}$, respectively) in our noise model.
Results with no simulated noise are provided at the x-origin of each plot to help differentiate the effects associated with the approximation depth, operand superposition, and gate error rates. The transparent, dashed, vertical lines at 0.2\% in the 1q-gate error rate plots and at 1.0\% in the 2q-gate error rate plots approximate the average reported performance of early IBM superconducting quantum computers \cite{cross2018ibm,murali2019full}. 

The construction and simulation of all circuits, noise models, and algorithms were implemented using Qiskit 0.31.0 and python 3.7.
\footnote{Our use of an outdated version of Qiskit was in an effort to maintain consistency in software versions for all data generated throughout the lifetime of this project. That said, we have regenerated samples of our data using Qiskit v1.0, and observed no statistically significant differences.} 
For all model calculations, circuits are decomposed in terms of the universal gate set of current IBM superconducting quantum computers (i.e., Id, X, RZ, SX, and CX-gates) \cite{cross2018ibm}. The operands' q-integer states were initialized using the reverse decomposition algorithm of Ref.\ \cite{2006shende} implemented in Qiskit. The 1q and 2q gate counts of the simulated circuits (excluding gates associated with initialization and measurement) are tabulated in Table~\ref{fig_gate_counts}.


We implement a success metric focused on the arithmetic output as opposed to state fidelity. When evaluating quantum circuits, one commonly uses success metrics like the state or Hellinger fidelity to gauge their success \cite{Jozsa1994FidelityFM}, as they offer a sense of how close the final state is to all aspects of the target output. This is particularly important for fault-tolerant algorithms, where the build-up of errors is a primary concern. However, because quantum arithmetic circuits containing multiple steps remains far beyond the current technology, a tomographic success metric (e.g. one focused solely on whether the correct arithmetic solution is obtained from the output) is of greater relevance when considering near-term and hybrid approaches. Further, having tabulated our results using both approaches, we found the tomographic results (in conjunction with their to-be-described error bar treatment) provided more direct insight. To this end, we implement a simple success metric using quantum state tomography, similar to that of Ref.~\cite{Methachawalit2020}. Each point in each figure marks the overall success rate in percentage (vertical axis) of 200 (or more) `instances' of the arithmetic operation on a random, unique choice of q-integers simulated at the given 1q-/2q-gate error rate, operand superposition, and AQFT approximation depth. For each instance, the simulated AQFA/AQFM circuit was repeated 2048 times (referred to as `shots'), and the outputs were tabulated. Instances were considered `successful' when the binary outputs with the highest frequency matched those anticipated from the input q-integers. In cases where multiple `correct' outputs were expected (due to superposed inputs), instances were deemed `unsuccessful' if any incorrect output possessed more counts than any one of the correct outputs.


Attached to each point, an associated pair of error bars are included to indicate the average ratio of correct to incorrect outputs observed for those results. In each instance, the minimum difference in counts between any single correct and incorrect output was recorded, and the standard deviation, $\sigma$, of these differences was calculated. The lower (upper) error bars therefore represent the number of `successful' (`unsuccessful') instances that would have been `unsuccessful' (`successful') within one $\sigma$, respectively. Consequently, narrow error bars reflect large differences between correct and incorrect output counts, while a wider error bar below/above the point reflects cases where many instances were on the verge of being evaluated as `unsuccessful'/`successful'.

\subsection*{Discussion}

With the details of the calculations specified, we now discuss our results. Beginning with the 1:1 addition results ((a) and (b) in Figs.~\ref{fig_sqfe},\ref{fig_qfe}), we see the outputs are largely insensitive to the regime of gate error rates we considered. Besides a small number of cases where the AQFT depth is one, all QFA operations (signed and unsigned) consistently produced correct results, and the lack of error bars demonstrates that, over all instances, the correct result counts were consistently at least one standard deviation higher than any other output. Even in the case of the $d=1$ calculations where the approximation is most severe and errors do occur (seen in how those points are below 100\%), the number of errors show little to no dependency on the gate error rates, indicating the errors are primarily the result of the approximation rendering the encoding too non-linear for some of the additions. The optimal choice of approximation depth appears to be $d>1$ for these cases, though any differences between the larger depths are not identifiable with the current approach. 


Moving to the 1:2 QFA results and 2:2 QFA results ((c), (d), (e), and (f) in Figs.~\ref{fig_sqfe},\ref{fig_qfe}), we immediately see greater sensitivity to the gate error rates in this regime. This agrees with the common observation that, for the near future, quantum applications that rely on or converge towards heavier output (that is, output corresponding to fewer, more highly-correlated binary strings in the computational basis) will be a driving design factor for early quantum applications, due to the significant noise observed in current hardware. (For instance, since each of the four correct outputs of the 2:2 results ideally possesses roughly $25\%$ of the probability density, fluctuations from noise have a far greater impact.) We also notice that while dependence on the 2q-error rate disappears at higher values than that of the 1q-gate error rate in these calculations, the impact of the 1q-gate error rate in current machines appears quite small (at least for QFA at $n=8$), compared to the larger number of errors corresponding to the current hardware's 2q-gate error rates. This supports the general expectation that 2q-gate fidelity is one current bottleneck for the performance of applications on modern quantum hardware. However, both here and in the 2:2 QFA results, we see a slower improvement of our result quality with reduced 1q-gate error compared to reduced 2q-gate error rates. If this trend continues at lower error rates and larger $n$, the 1q-gate errors could overcome 2q-gate errors as a limiting factor for the QFA at high $n$. Determining whether this trend continues to hold will require further simulation and study. 

Unlike the 1:1 QFA results, both the 1:2 and 2:2 QFA results provide much more insight regarding the optimal approximation depth. Even just considering only the reduction in error garnered by the AQFT in the presence of either 1q- or 2q-gate error, we generally see at least similar levels of performance to the full QFT for depths near $d=\log_2 n$, and frequently a slight improvement. One anticipates greater benefit on a real machine, where the AQFT's reduction in gate count will also mitigate errors associated with reset, measurement, qubit-connectivity, and other sources of noise. As surmised in \cite{1996barenco}, the quality of results do appear to be somewhat higher near $\log_2 n=3$. However, we do observe some variation in the optimal approximation depth between clusters in and between all four plots. Depths 2, 3 and 4 are the most common optima, though generally by relatively small margins. Even beyond the noise-free simulations, the AQFT at depth 1 generally produces significantly lower quality results, supporting how, even at $n=8$, going to too deep an approximation depth eventually results in a rapid drop in performance. 

If we carefully compare the signed and unsigned QFA results, one notices the signed results are often significantly more accurate than the unsigned results, sometimes by margins as large as 40 points. Given the use of identical code versions, hardware, and error rates when generating the data for each, we would ostensibly expect such differences to remain nominal in the converged results. In essence, the only material differences in the two calculations are the inclusion of the two Toffoli gates shown in the middle panel of Fig.~\ref{fig_qadd_gate1} that essentially flip whether the phase-shift will be additive or subtractive based on the q-integer addends' sign breakdown. The inclusion of the extra gates (in the presence of non-zero gate error rates) would generally be expected to marginally reduce the result's accuracy, yet here their inclusion seems to substantially improve performance. The increased performance appears fairly uniform across gate error rates and AQFT depths.
Further testing reveals the improved performance in the signed results persists even if one only considers positive integer states, and only disappears with the removal of the aforementioned gates from the circuit. Why their inclusion has so noticeable an impact on performance, let alone one of improvement, remains unclear. We hope further analysis using a more robust model regime may help elucidate these apparent differences in a future study.


We include the QFM results from Ref.~\cite{basili2021performance} in Fig.~\ref{fig_qfm} as a point of comparison with the new QFA results. As anticipated, the greater circuit depth makes 2q-gate error even more significant, despite the much smaller register size. See the aforementioned reference for a complete analysis of the results. Whether the inclusion of sign to QFM results in a similar increase in performance as that observed in the QFA results of this work remains a point of study.

Finally, we see a common shape throughout both past and present results with changing gate error rate, involving a roughly linear trend of the success rate between asymptotic tails on the left/right side, roughly approaching 100\%/0\% success rates, respectively. This is common to sampling over Gaussian distributions, and reflects that the impact of gate errors on the arithmetic success rate follows a similar distribution. The width of the linear region reflects the width of a `plateau' in the statistical distribution, meaning a smaller interval of linearity corresponds to a sharper threshold for success.
We also notice in (c) and (e) of the QFA results that this region grows narrower as we increase the number of superposed operand states. If we assume this trend continues, one expects that the gate fidelity required for highly entangled arithmetic will be a relatively sharp threshold. Like the 2-bit results of quantum addition seen in Ref.~\cite{Methachawalit2020}, with sufficiently high gate error rates and order of superposition, we consistently observe results with an accuracy approaching 0\%. This reflects that the length of the circuit and poor gate fidelity create too much noise to extract the solution.

\begin{table}[t]
\caption{Arithmetic Circuit Gate Counts. Note signed QFA adds an additional four 1q-gates and ten 2q-gates.}
\begin{center}
\begin{tabular}{|c||c|c|c|c|c||c|c|c|}
\cline{2-9}
\multicolumn{1}{c|}{}  & \multicolumn{5}{|c|}{\textbf{QFA ($n=8$)}} & \multicolumn{3}{|c|}{\textbf{QFM ($n=4$)}} \\
\hline
\textbf{$d^{*}$} & 1 & 2 &3& 4 & $7^{**}$ & 1 & 2 & $3^{**}$  \\
\hline
\hline
\textbf{1q} & 163 & 199 & 229 &253 & 289 & 1032 & 1248 & 1464 \\
\hline
\textbf{2q} & 98 & 122 & 142 & 158 & 182 & 744 & 936 & 1128 \\
\hline
\multicolumn{9}{l}{$^*$AQFT approximation depth}\\
\multicolumn{9}{l}{$^{**}$This depth is equivalent to the full QFT for the given $n$.}

\end{tabular}
\label{fig_gate_counts}
\end{center}
\end{table}

\section{Conclusions and Future Works}
\label{sec_summary}
In this work, we simulated the dependency of signed and unsigned QFA and unsigned QFM on one- and two-qubit gate error rates, approximation depth in the QFT, and operand's order of superposition. We utilized noise models to isolate sources of error and compared the results to perfect, noise-free simulations to differentiate noise inherent to the approximate QFT. In a broader sense, this work reflects an initial effort to better assess the practical challenges that must be overcome before a quantum numerical paradigm can be realistically explored.  

The use of the AQFT proved very effective in both our signed and unsigned QFA results, often generating little to no visible negative impacts, even in error-free simulation. In the presence of noise, results with the AQFT near the optimal approximation almost always produced higher quality results than those of the full QFT. However, even at low qubit count, $n$, we observed how going to too low an approximation depth has a severe negative impact on our result quality. Further, while a depth of $d=\log_2 n$ served as a reasonable heuristic, we observed significant variation on the optimal approximation depth between the orders of superposition, and even between changes in the gate error rates alone.

Our results support the expectation that 2q-gate error rates currently pose the greater obstacle for quantum arithmetic, though we also observed that our success rates improved much faster as the 2q-gate error rate falls than it does as the 1q-gate error rate falls. Whether this suggests 1q-gate errors will surpass 2q-gate errors as a limiting factor at high $n$ for the discussed arithmetic approaches remains possible but inconclusive.

As anticipated, we observed significant drops in performance as we increased the order of superposition, and that lower error rates significantly reduced this drop in performance. For instance, at the optimal depth, increasing 1:2 addition to 2:2 addition at $n=8$ resulted in over a 50\% drop in accuracy at the current 2q-gate error rate of recent IBM hardware, but only a roughly 3\% drop when simulated with the improved 2q-gate error rate of 0.7\%. Consequently, the possibility of a quantum numerical computational paradigm will rely heavily on improving current hardware.

Finally, we observed a substantially better overall performance in the results of signed QFA over unsigned QFA. This difference appears to depend solely on the inclusion of the Toffoli gates shown in the middle panel of Fig.~\ref{fig_qadd_gate1} in the simulated circuit. The source of this difference remains unknown, and demands further study. Confirming this difference on a physical quantum device and forming a theoretical model to explain it would be particularly informative. 

This work sheds light on the current challenges and hardware limitations that must be resolved before a quantum numerical computational paradigm becomes a more realistic possibility. Our results provide insight on what quantum arithmetic will be feasible, and we have observed greater complexity in determining the optimal AQFT depth than we initially anticipated. 

One natural step forward is to continue our investigation of isolating different sources of error (read-out, decoherence time, hardware connectivity, etc.) and their interdependency on order of superposition, as well as to what degree the AQFT mitigates them. Another natural option would be to extend our study to signed and/or unsigned division, exponentiation, and floating-point operations. Noise mitigation techniques, such as zero-noise extrapolation and symmetry checks, would provide useful insight relevant to quantum arithmetic in general.
Looking farther into the future, greater variation on how superposed states are entangled may also be informative for arithmetic on fault-tolerant quantum systems. Simulations that isolate other sources of error, such as thermal relaxation, qubit measurement, their combination with 1q-/2q- gate errors, as well as the impact of error mitigation and extrapolation from integers to real numbers, would offer a much more complete understanding of how these elements impact arithmetic with the QFT and AQFT. Extending the study to larger $n$ would reveal whether the trends observed here continue or diverge, and possibly other characteristics that were suppressed at the smaller register sizes considered here. This is particularly true for QFM, which was performed only for $N=4$. 

Changing how sign is encoded or its implementation on arithmetic circuits may shed light on the improved performance observed for signed over unsigned addition in this study. Designing, implementing, and simulating signed QFM may similarly offer insight for the discrepancy, as well as
reveal critical insight into current and new quantum algorithms, such as those for weighted-sum problems. Moreover, they will lay necessary foundation as future improvements in technology and techniques continue to make a quantum numerical computational paradigm more feasible.



\section*{Acknowledgments}
We acknowledge fruitful discussions with Soham Pal and Weijie Du. We acknowledge the use of IBM Quantum services for this work. The views expressed are those of the authors, and do not reflect the official policy or views of IBM or the IBM Quantum team. This article is a revised and expanded version of a paper entitled "Performance Evaluations of Noisy Approximate Quantum Fourier Arithmetic'', which was presented at the 2022 IEEE International Parallel and Distributed Processing Symposium Workshops (IPDPSW) in Lyon, France \cite{basili2021performance}. This research was funded by the US Department of Energy (DOE) under Grant Nos. DE-SC0023495 (SciDAC5/NUCLEI) and DE-SC0023707 (Quantum Horizons/NuHaQ) through the Office of Nuclear Physics Quantum Horizons program for the "{\bf Nu}clei and {\bf Ha}drons with {\bf Q}uantum computers ({\bf NuHaQ})” project), and by NSF Grant No. 2435255 (\textbf{NQVL-QSTD: Q-BLUE}).


\bibliographystyle{unsrtnat}
\bibliography{references}

\begin{thebibliography}{39}
\providecommand{\natexlab}[1]{#1}
\providecommand{\url}[1]{\texttt{#1}}
\expandafter\ifx\csname urlstyle\endcsname\relax
  \providecommand{\doi}[1]{doi: #1}\else
  \providecommand{\doi}{doi: \begingroup \urlstyle{rm}\Url}\fi

\bibitem[Shor(1997)]{shor1997}
Peter~W. Shor.
\newblock Polynomial-time algorithms for prime factorization and discrete logarithms on a quantum computer.
\newblock \emph{SIAM Journal on Computing}, 26\penalty0 (5):\penalty0 1484–1509, Oct 1997.
\newblock ISSN 1095-7111.
\newblock \doi{10.1137/s0097539795293172}.
\newblock URL \url{http://dx.doi.org/10.1137/S0097539795293172}.

\bibitem[Preskill(2018)]{Preskill2018quantumcomputingin}
John Preskill.
\newblock Quantum {C}omputing in the {NISQ} era and beyond.
\newblock \emph{{Quantum}}, 2:\penalty0 79, August 2018.
\newblock ISSN 2521-327X.
\newblock \doi{10.22331/q-2018-08-06-79}.
\newblock URL \url{https://doi.org/10.22331/q-2018-08-06-79}.

\bibitem[Kopczyk(2018)]{kopczyk2018quantum}
Dawid Kopczyk.
\newblock Quantum machine learning for data scientists, 2018.

\bibitem[Chakrabarti and Sur-Kolay(2008)]{chakrabarti2008}
Amlan Chakrabarti and Susmita Sur-Kolay.
\newblock Designing quantum adder circuits and evaluating their error performance.
\newblock In \emph{2008 International Conference on Electronic Design}, pages 1--6, 2008.
\newblock \doi{10.1109/ICED.2008.4786689}.

\bibitem[Methachawalit and Chongstitvatana(2020)]{Methachawalit2020}
Wiphoo Methachawalit and Prabhas Chongstitvatana.
\newblock Adder circuit on ibm universal quantum computers.
\newblock In \emph{2020 17th International Conference on Electrical Engineering/Electronics, Computer, Telecommunications and Information Technology (ECTI-CON)}, pages 92--95, 2020.
\newblock \doi{10.1109/ECTI-CON49241.2020.9158064}.

\bibitem[Leymann and Barzen(2020)]{Leymann_2020}
Frank Leymann and Johanna Barzen.
\newblock The bitter truth about gate-based quantum algorithms in the {NISQ} era.
\newblock \emph{Quantum Science and Technology}, 5\penalty0 (4):\penalty0 044007, sep 2020.
\newblock \doi{10.1088/2058-9565/abae7d}.
\newblock URL \url{https://doi.org/10.1088/2058-9565/abae7d}.

\bibitem[Basili et~al.(2022)Basili, Qian, Tang, Castellino, Eshaghian-Wilner, Khokhar, Luecke, and Vary]{basili2021performance}
Robert Basili, Wenyang Qian, Shuo Tang, Austin Castellino, Mary Eshaghian-Wilner, Ashfaq Khokhar, Glenn Luecke, and James~P. Vary.
\newblock Performance evaluations of noisy approximate quantum fourier arithmetic.
\newblock In \emph{2022 IEEE International Parallel and Distributed Processing Symposium Workshops (IPDPSW)}, pages 435--444, 2022.
\newblock \doi{10.1109/IPDPSW55747.2022.00081}.

\bibitem[Draper(2000)]{draper2000addition}
Thomas~G Draper.
\newblock Addition on a quantum computer.
\newblock \emph{arXiv preprint quant-ph/0008033}, 2000.

\bibitem[{\'{A}}lvarez-S{\'{a}}nchez et~al.(2008){\'{A}}lvarez-S{\'{a}}nchez, {\'{A}}lvarez-Bravo, and Nieto]{2008varez}
J~J {\'{A}}lvarez-S{\'{a}}nchez, J~V {\'{A}}lvarez-Bravo, and L~M Nieto.
\newblock A quantum architecture for multiplying signed integers.
\newblock 128:\penalty0 012013, aug 2008.
\newblock \doi{10.1088/1742-6596/128/1/012013}.
\newblock URL \url{https://doi.org/10.1088/1742-6596/128/1/012013}.

\bibitem[Ruiz-Perez and Garcia-Escartin(2017)]{ruiz2017quantum}
Lidia Ruiz-Perez and Juan~Carlos Garcia-Escartin.
\newblock Quantum arithmetic with the quantum fourier transform.
\newblock \emph{Quantum Information Processing}, 16\penalty0 (6):\penalty0 152, 2017.

\bibitem[Babu(2017)]{babu2017cost}
Hafiz Md~Hasan Babu.
\newblock Cost-efficient design of a quantum multiplier--accumulator unit.
\newblock \emph{Quantum Information Processing}, 16\penalty0 (1):\penalty0 1--38, 2017.

\bibitem[Şahin(2020)]{2020sahin}
Engin Şahin.
\newblock Quantum arithmetic operations based on quantum fourier transform on signed integers.
\newblock \emph{International Journal of Quantum Information}, 18\penalty0 (06):\penalty0 2050035, Sep 2020.
\newblock ISSN 1793-6918.
\newblock \doi{10.1142/s0219749920500355}.
\newblock URL \url{http://dx.doi.org/10.1142/S0219749920500355}.

\bibitem[Kurt et~al.(2023)Kurt, Kaltehei, Kuzu, Gencen, and Cakmak]{kurt2023qft}
Murat Kurt, Ayda Kaltehei, Ahmet Kuzu, Azmi Gencen, and Selcuk Cakmak.
\newblock Qft based quantum arithmetic logic unit on ibm quantum computer, 2023.

\bibitem[Kurt et~al.(2024)Kurt, Kaltehei, Gençten, and Çakmak]{kurt2024scalable}
Murat Kurt, Ayda Kaltehei, Azmi Gençten, and Selçuk Çakmak.
\newblock Scalable quantum circuit design for qft-based arithmetic, 2024.

\bibitem[Yuan et~al.(2023)Yuan, Wang, Wang, Chen, Dou, Wu, and Guo]{yuan2023improved}
Yewei Yuan, Chao Wang, Bei Wang, Zhao-Yun Chen, Meng-Han Dou, Yu-Chun Wu, and Guo-Ping Guo.
\newblock An improved qft-based quantum comparator and extended modular arithmetic using one ancilla qubit, 2023.

\bibitem[Crimmins(2023)]{crimmins2023efficient}
Aden~L Crimmins.
\newblock Efficient quantum integer multiplication in the quantum fourier transform domain.
\newblock Master's thesis, Wright State University, 2023.

\bibitem[Ramezani et~al.(2023)Ramezani, Nikaeen, Farman, Ashrafi, and Bahrampour]{ramezani2023quantum}
Mehdi Ramezani, Morteza Nikaeen, Farnaz Farman, Seyed~Mahmoud Ashrafi, and Alireza Bahrampour.
\newblock Quantum multiplication algorithm based on the convolution theorem.
\newblock \emph{Physical Review A}, 108\penalty0 (5):\penalty0 052405, 2023.
\newblock \doi{10.1103/PhysRevA.108.052405}.

\bibitem[Wang et~al.(2024{\natexlab{a}})Wang, Li, Lee, Deb, Lim, and Chattopadhyay]{wang2024comprehensive}
Siyi Wang, Xiufan Li, Wei Jie~Bryan Lee, Suman Deb, Eugene Lim, and Anupam Chattopadhyay.
\newblock A comprehensive study of quantum arithmetic circuits, 2024{\natexlab{a}}.

\bibitem[Chen et~al.(2024)Chen, Song, Zhou, and Wen]{chen2024quantum}
Zhao-Yun Chen, Cheng-Qian Song, Yu-Cheng Zhou, and Jin-Ming Wen.
\newblock Quantum arithmetic based on quantum signal processing, 2024.

\bibitem[Joshi et~al.(2023)Joshi, Chandarana, Katragadda, Tejeswini, Vallam, and Manjrekar]{joshi2023quantum}
Aravind Joshi, Pranav Chandarana, Chaitanya Katragadda, Vishnu~P Tejeswini, Rohith~D Vallam, and Madhavan Manjrekar.
\newblock Quantum circuit optimization of arithmetic circuits using zx calculus, 2023.

\bibitem[Wang et~al.(2024{\natexlab{b}})Wang, Deb, Mondal, and Chattopadhyay]{wang2024optimal}
Siyi Wang, Suman Deb, Ankit Mondal, and Anupam Chattopadhyay.
\newblock Optimal toffoli-depth quantum adder, 2024{\natexlab{b}}.

\bibitem[Babu and Kumar(2022)]{babu2022higher}
H~Thameem~Ansari Babu and R~Sakthivel Kumar.
\newblock A higher radix architecture for quantum carry-lookahead adder.
\newblock \emph{Scientific Reports}, 12\penalty0 (1):\penalty0 12372, 2022.
\newblock \doi{10.1038/s41598-022-16179-w}.

\bibitem[Gidney(2018)]{gidney2018halving}
Craig Gidney.
\newblock Halving the cost of quantum addition.
\newblock \emph{Quantum}, 2:\penalty0 74, 2018.
\newblock \doi{10.22331/q-2018-06-18-74}.

\bibitem[Wang et~al.(2024{\natexlab{c}})Wang, Lim, and Chattopadhyay]{wang2024boosting}
Siyi Wang, Eugene Lim, and Anupam Chattopadhyay.
\newblock Boosting the efficiency of quantum divider through effective design space exploration.
\newblock In \emph{2024 IEEE International Symposium on Circuits and Systems (ISCAS)}, pages 1--5. IEEE, 2024{\natexlab{c}}.
\newblock \doi{10.1109/ISCAS57851.2024.10590686}.

\bibitem[Gidney(2024)]{gidney2024fast}
Craig Gidney.
\newblock Fast quantum integer multiplication with zero ancillas, 2024.

\bibitem[Zhan(2023)]{zhan2023quantum}
Junpeng Zhan.
\newblock Quantum multiplier based on exponent adder, 2023.

\bibitem[Cao et~al.(2019)Cao, Liu, Zhang, Zhang, and Zhang]{cao2019designs}
Guang-Can Cao, Bin-Ze Liu, Wei-Wei Zhang, Jin-Yu Zhang, and Ru~Zhang.
\newblock Designs of the divider and special multiplier optimizing t and cnot gates.
\newblock \emph{Quantum Engineering}, 1\penalty0 (2):\penalty0 e16, 2019.
\newblock \doi{10.1002/que2.16}.

\bibitem[Yang et~al.(2022)Yang, Zhou, Wang, Xu, and Sun]{yang2022circuit}
Zhao-Xu Yang, Ran Zhou, Zhi-Ping Wang, Kun Xu, and Xiao-Hui Sun.
\newblock The circuit design and optimization of quantum multiplier and divider.
\newblock \emph{Science China Physics, Mechanics \& Astronomy}, 65\penalty0 (6):\penalty0 260311, 2022.
\newblock \doi{10.1007/s11433-021-1838-8}.

\bibitem[Gouzien and Saeedi(2020)]{gouzien2020improving}
{\'E}tienne Gouzien and Mehdi Saeedi.
\newblock Improving the number of t gates and their spread in integer multipliers on quantum computing.
\newblock \emph{Physical Review A}, 102\penalty0 (4):\penalty0 042417, 2020.
\newblock \doi{10.1103/PhysRevA.102.042417}.

\bibitem[Luongo et~al.(2025)Luongo, Narasimhachar, and Sireesh]{luongo2025optimized}
Alessandro Luongo, Varun Narasimhachar, and Adithya Sireesh.
\newblock Optimized circuits for windowed modular arithmetic with applications to quantum attacks against rsa, 2025.

\bibitem[Ao et~al.(2023)Ao, Yang, and Wang]{ao2023implementation}
Si-Yuan Ao, Ming-Zhong Yang, and He-Liang Wang.
\newblock The implementation of the enhanced quantum floating-point adder.
\newblock \emph{Modern Physics Letters A}, 38\penalty0 (01):\penalty0 2250211, 2023.

\bibitem[H{\"a}ner et~al.(2018)H{\"a}ner, Roetteler, and Svore]{haner2018quantum}
Thomas H{\"a}ner, Martin Roetteler, and Krysta~M Svore.
\newblock Quantum circuits for floating-point arithmetic.
\newblock In \emph{Reversible Computation: 10th International Conference, RC 2018, Leicester, UK, September 13-14, 2018, Proceedings 10}, pages 33--47. Springer, 2018.
\newblock \doi{10.1007/978-3-319-99498-7_3}.

\bibitem[Li(2024)]{li2024feasible}
Junxu Li.
\newblock A feasible design of elementary quantum arithmetic logic units for near-term quantum computers, 2024.

\bibitem[Fillinger(2013)]{fillinger2013data}
Maximilian Fillinger.
\newblock Data structures in classical and quantum computing, 2013.

\bibitem[Cross(2018)]{cross2018ibm}
Andrew Cross.
\newblock The ibm q experience and qiskit open-source quantum computing software.
\newblock In \emph{APS March Meeting Abstracts}, volume 2018, pages L58--003, 2018.

\bibitem[Barenco et~al.(1996)Barenco, Ekert, Suominen, and Törmä]{1996barenco}
Adriano Barenco, Artur Ekert, Kalle-Antti Suominen, and Päivi Törmä.
\newblock Approximate quantum fourier transform and decoherence.
\newblock \emph{Physical Review A}, 54\penalty0 (1):\penalty0 139–146, Jul 1996.
\newblock ISSN 1094-1622.
\newblock \doi{10.1103/physreva.54.139}.
\newblock URL \url{http://dx.doi.org/10.1103/PhysRevA.54.139}.

\bibitem[Murali et~al.(2019)Murali, Linke, Martonosi, Abhari, Nguyen, and Alderete]{murali2019full}
Prakash Murali, Norbert~Matthias Linke, Margaret Martonosi, Ali~Javadi Abhari, Nhung~Hong Nguyen, and Cinthia~Huerta Alderete.
\newblock Full-stack, real-system quantum computer studies: Architectural comparisons and design insights.
\newblock In \emph{2019 ACM/IEEE 46th Annual International Symposium on Computer Architecture (ISCA)}, pages 527--540. IEEE, 2019.

\bibitem[Shende et~al.(2006)Shende, Bullock, and Markov]{2006shende}
V.V. Shende, S.S. Bullock, and I.L. Markov.
\newblock Synthesis of quantum-logic circuits.
\newblock \emph{IEEE Transactions on Computer-Aided Design of Integrated Circuits and Systems}, 25\penalty0 (6):\penalty0 1000–1010, Jun 2006.
\newblock ISSN 1937-4151.
\newblock \doi{10.1109/tcad.2005.855930}.
\newblock URL \url{http://dx.doi.org/10.1109/TCAD.2005.855930}.

\bibitem[Jozsa(1994)]{Jozsa1994FidelityFM}
Richard Jozsa.
\newblock Fidelity for mixed quantum states.
\newblock \emph{Journal of Modern Optics}, 41:\penalty0 2315--2323, 1994.

\end{thebibliography}

\end{document}